# Pragmatism in industrial modelling, applied to "ladle lifetime in the steel industry"

By Stein Tore Johansen[1*], Bjørn Tore Løvfall[1], Tamara Rodriguez Duran[2] and Josip Zoric[1]

[1]     SINTEF; Stein Tore Johansen, Bjørn Tore Løvfall, Josip Zoric

[2]     Sidenor; Tamara Rodriguez Duran

*     Correspondence: Stein Tore Johansen; stein.t.johansen@sintef.no

*Abstract*

A methodology for building pragmatic physics based models (Zoric et al., 2015b) is here adapted to a use-case in the steel industry. The challenge is to predict the erosion of steel ladle linings, such that the model can support operators to decide if the lade lining can be used one more time or not. If the ladle has too thin lining 140 tons of hot liquid steel may escape out of the ladle, with huge consequences for workers and plant. The development was done with a very small core team (two developers), which is typical for many industrial developments. The adopted workflow for the development, challenges that were faced, and some model results are presented. One key learning is that development of models should allow time for maturing the process understanding, and time should be given for many iterations by "questions-responses and actions" at the various levels in the model development. The good interactions between development team and industry case owner is an important success factor. In this case the results of using the PPBM (Pragmatism in physics-based modelling) were good thanks to very successful interaction between development team and industry case owner.

Combining or extending the model with use of ML methods and cognition-related methods, such as knowledge graphs and self-adaptive algorithms is discussed.

**Introduction**

In many industrial processes complex physics and chemistry are involved. In addition, the observability (Wikipedia, 2022a) may be very poor and accordingly process control becomes hard. In high temperature processes, including metallurgical industry, we find many representative cases. There may be some available sensors in specific cases, but due to excessive purchase and maintenance costs these are out of the question. Physics based models can be a remedy and predict internal states of the process, if based on a realistic set of assumptions and simplifications. Still, this is the only way to learn about the inner workings of the process. The amount of data in this type of processes is limited, and some of the data may have significant issues as a result of operational challenges. The operations are not automated, and sometimes fast decisions must be made without time to check if the change in operation impacts the data collection. As a result of these challenges we have been developing a generic methodology (Johansen, S. T. et al., 2017; Johansen and Ringdalen, 2018; Zoric et al., 2015b, 2015a) framed as " Pragmatism in industrial modelling", for the development of industry applicable physics-based models.

Earlier work on pragmatism (Johansen, S. T. et al., 2017; Zoric et al., 2015a, 2015b) addressed solving various industrial tasks and problems. Here the best approaches where proposed and discussed, by assuming no limitations in the human expertise and resources. A generic development scheme for our pragmatic development approach is shown in Figure 1.

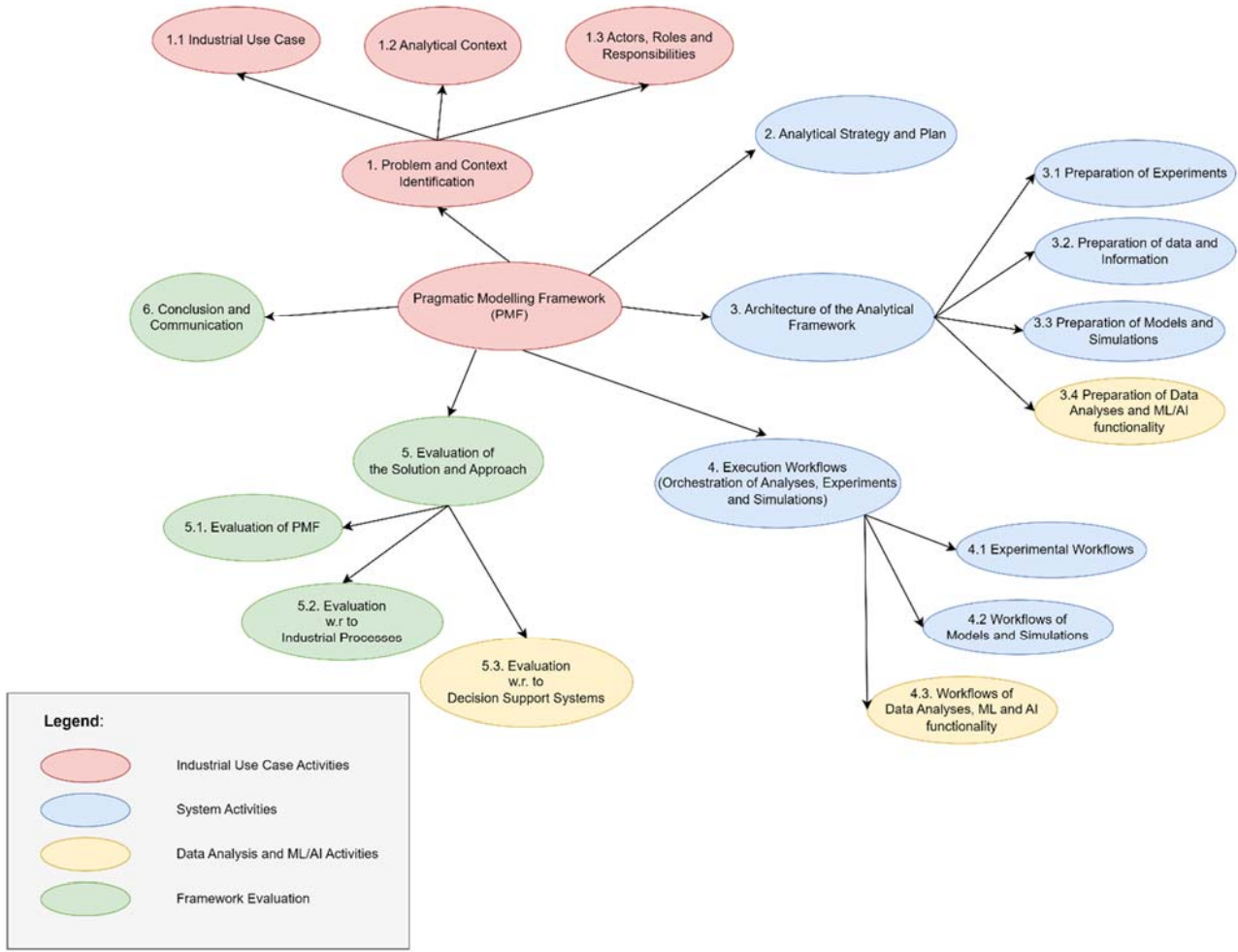

*Figure 1 Proposed workflow for a pragmatism based development (Zoric et al., 2015b). Arrows indicate sub use cases and related activities that have to be realized.*

In this industrial process the use of the pragmatic modelling paradigm will hopefully: (1) handle complexity of a heterogeneous and hybrid modelling approach, (2) structure the exchange of data and information between sub-models (often of different abstraction levels), (3) structure the modelling and analytic workflows and (4) give good interface places to include some ML/AI prediction tools and (5) connect these tools to the relevant decision support tools and processes. Inclusion of modelling and simulation frameworks in industrial processes and industrial decision support systems requires significant structure and standardization, with which we hope to contribute to by this work.

Pragmatic modelling starts always with a given industrial application, by defining an industrial use case (i.e. problem to solve) (Wikipedia, 2022b). The pragmatic model is the simplest model which can give fast, and sufficiently good answers. There could be a short step from a pragmatic model to online process control and operation support tools. A pragmatic model starts with the simplest possible model which has a value for the user. The pragmatic workflow has main steps, as seen in Figure 1:

1. Problem and Context Identification
2. Analytical Strategy and Plan
3. Architecture of the Analytical Framework
4. Execution (Orchestration of Analyses, Simulations and Experiments)
5. Evaluation of the Solution
6. Conclusion and Communication

A key element in this work is the appointment of the system architect team ((Zoric et al., 2015a, 2015b),(Wikipedia, 2022c),(Wikipedia, 2022d)), as the capabilities of this team will be a critical factor for a successful outcome of the work.

The methodology, which the system architects team has to orchestrate is not limited to any specific techniques (steps 2-4), and the palette of tools may contain elements that involve mathematical methods, such as statistical methods, singular value decomposition (Wikipedia, 2022e) and reduced order methods (Wikipedia, 2022f) but also numerical continuum physics, numerical particle physics, including molecular and quantum mechanics. In practice the system architect team will learn to apply and orchestrate the methods that are at hand for the development team (methods applicable to the reality of the industrial process, and related decision support systems and routines). It should be noticed that the methods of machine learning and artificial intelligence are already included in the above-mentioned methods. The pragmatism-based methodology will use available sensor data and access the validity of the data. However, development of new sensors, even if critical, is not dealt with by the methodology.

In the present paper we aim to present a simplified pragmatism-based approach for the development of a prediction model steel ladle refractory erosion and lifetime. A particular challenge is that the total development team is small (2-3 people) and multiple trade-offs must be made to develop a useful model in a limited time.

### Context of the COGNITWIN project

This work, as a part of the Horizon 2020 project (COGNITWIN, 2022), is aiming to accelerate the digital transformation and introduce Industry 4.0 to the European process industries. The project is focused around 6 industrial pilots, ranging from aluminium production, silicon production, steel production to engineering. We will here address the pilot for the Sidenor[1] steel company, where we analyse how to increase the ladle refractory lifetime, and how the digital twin concept can contribute.

*Sidenor ladle case description*

The steel production in the melting shop process is based on three main steps. The first one lies in obtaining the liquid steel by melting iron ore (blast furnace, BF) or scrap (electrical arc furnace, EAF or induction furnace). The second step (Secondary Metallurgy, SM) is necessary for refining the liquid steel, and the last one solidifies the steel during ingot or continuous casting processes.

Typically, a ladle can contain from tens to hundreds of liquid steel tons (Figure 2). Most ladles have installed a porous plug at the bottom. Gas (Ar of N) injected through the plug is responsible for the liquid steel stirring. The rising flow of the liquid steel promotes the inclusion decantation from the steel to the slag and homogenizes the temperature and chemical composition.

The main objective of the SM is to obtain the correct chemical composition and to ensure enough temperature for the casting process. In addition, there are several important tasks which must be complete during the SM, as for example inclusion and gases removal. In order to reach these objectives, Sidenor has a SM mill consisting of two Ladle furnaces (LF) and a Vacuum Degasser (VD). Each of the LFs have three electrodes, which are responsible of heating the slag, steel, and ferro-additions. The ladle contains liquid steel and slag for all the production process, from the EAF to the end of the casting process. The liquid steel has a temperature of around 1850 K in the ladle, and it is covered with slag. The slag avoids the contact between the liquid steel and the atmosphere, has lower density than steel and consists basically of lime and oxide elements. The slag conditioning can be improved during the SM by adding slag-formers.

---

[1] https://www.sidenor.com/en/

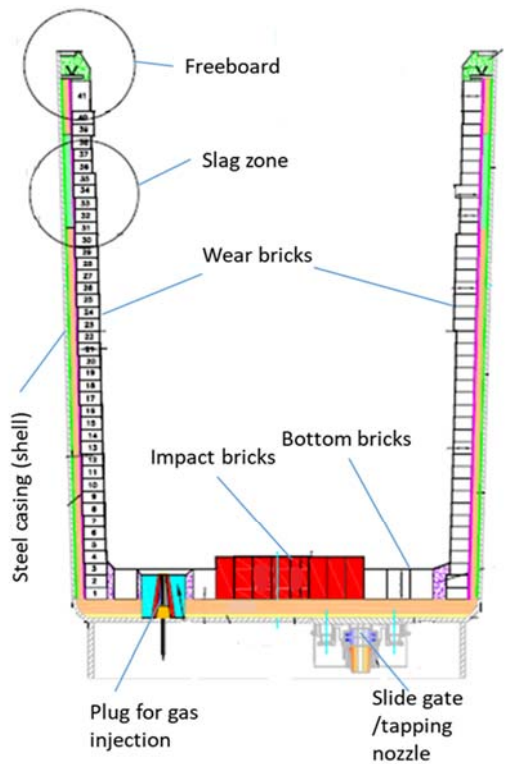 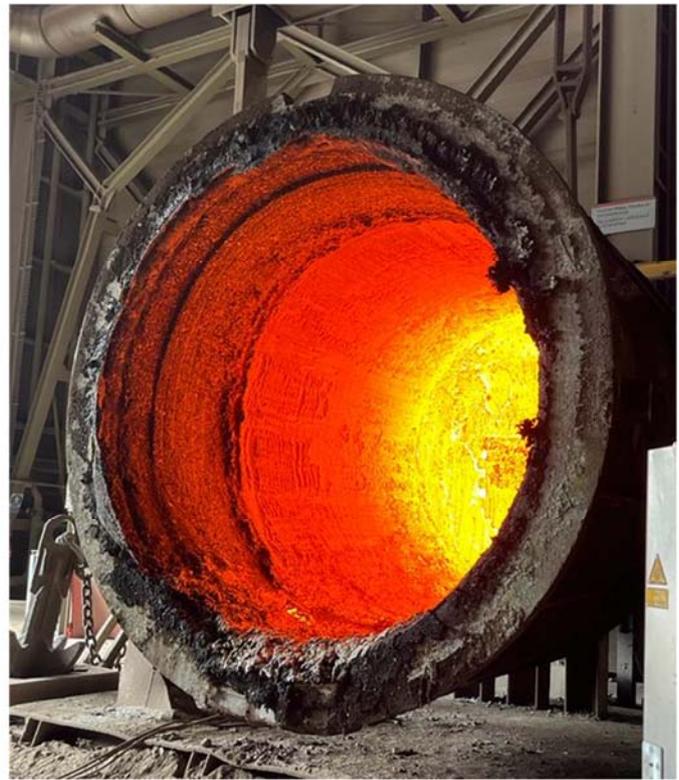

*Figure 2 Left: Sketch of typical steel ladle, with wear refractory bricks, permanent lining (between wear bricks and steel casing), steel casing, bottom bricks, bottom plug for gas blowing and tapping hole (nozzle) with slide gate. Right: Hot ladle that has been in use and is waiting for the next heat. Maximum steel capacity is around 150 tons.*

In order to handle the liquid steel and slag with such high temperature, the ladle is built with a strong outer steel shell, the inside of which is covered *with layers of insulating materials (refractory). The refractory is made of ceramic and its most important properties are:*

  i.   ability to handle high temperature
 ii.   favourable thermal properties
iii.   high resistance against erosion when in contact with steel and slag

The inner layer of refractory bricks, which are in contact with the liquid steel, is eroded by the interaction with the hot metal and the slag. The bricks are eroded away by each heat, and after several heats they are so eroded that it is not safe to use the ladle one more time. The refractory is visually inspected after each heat and depending on its state, the ladle may be used one more heat, put aside for repair or demolished. In case of repair, the upper bricks of the ladle, which are more eroded, will be replaced by new ones. Once the ladle is repaired, it is taken back into production. Later, based on continuing visual inspection, the ladle may be deemed ready for demolition. In this case the entire inner lining is removed and relined with new bricks.

One important goal for Sidenor is to reduce the refractory costs by finding new methods for extending the refractory life. One of the key points is to use the same ladle during more heats without compromising the safety, but another important issue is to understand better the mechanism that drives the refractory erosion, avoid as much as possible the bad working practices and in such a way enlarge the working life.

### *Target for the pragmatic model development*

The main goal of our pragmatic modelling approach is to develop a model where the results can help deciding whether the ladle could be used one more time safely. The model should exploit both historic and current production data. The model increases the knowledge of the operators. It could contribute to related digital twins also in semantic and cognitive aspects.

In addition, the model should give information about which parameters dominate the ladle refractory erosion and give tips about which precautions may be taken to extend the refractory lifetime.

### **Physics-based pragmatic model**

We will now investigate the recommended workflow for the development of the above-mentioned pragmatic model. The generic workflow could be applied to any development of physics-based models. A particular ambition with this work is to develop a hybrid model that can base predictions on any combinations of direct use of data, indirect use of data and the physics-based model. However, for the physics-based model the data is crucial to tune the model. The justification of tuning is that we are dealing with an extremely complex process, containing multiple levels of uncertainty. As part of the overall complexity many critical physical data are unknown or have changed due to aging.

It was decided to frame the model as PPBM (Pragmatism in physics-based modelling). Referring to the pragmatism steps 1-6 above, we first set out to establish the development team, comprising of mainly two developers. This step is preparatory, and the team was selected based on experience and skills. In addition, the contributions from the industry were crucial to understand the case and provide relevant data.

The following text describes the PPBM development in six steps, as illustrated in Figure 1.

### *Pragmatism Step 1 Problem and Context Identification*

Step 1 aims at describing accurately the purpose of the model and the quantitative output data the model shall produce, including time constraints and accuracy requirements. To facilitate this step we employ the user stories described from the perspective of ladle operators at Sidenor / industry case owner. The main user, who is expected to benefit from the model results, places the model in the industrial process perspective and defines its role and contribution in the industry process perspective. We discuss our experiences in solving the challenges met during that collaboration, which we experienced as very demanding. The actors and entities participating in the overall case workflow are shown in Figure 3. The step *(1) Definition Accepted* should have been finished latest after 6 months. However, the problem definition and context was continuously challenged all the way without any formal requests to change the definition. In addition to developing the physics-based model the objective included to develop the model in such a way that it can be used in different hybrid approaches, combining data and physics-based models. The hybrid approach involving using data to calibrate the model is included here, while development of models that explore the combination of the physics-based model and all additional available data is outside the scope of this development. However, continuous interaction between the machine learning team (MLT) and the physics-based team had to be taken care of. Important decision gates during the model development process are illustrated in Table 1.

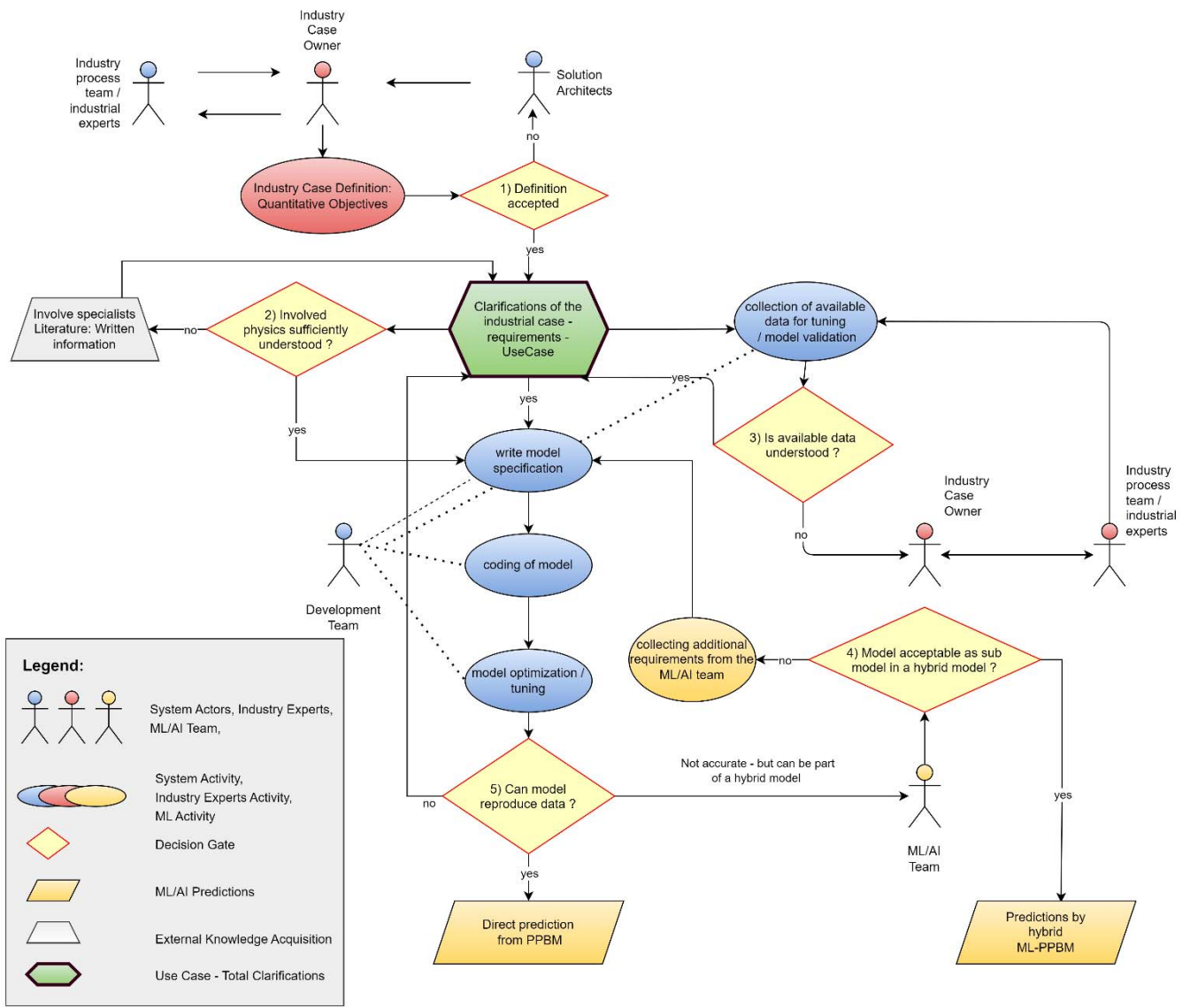

*Figure 3 Development steps for the Pragmatism in physics-based modelling (PPBM)*

*User story*

The main result of this phase, the user story can be summarized as follows. A model of the SM that can predict the average refractory loss for a given use of the ladle, and the accumulated loss over the lifetime of the ladle. The expected input parameters are the amount of steel, slag and additives added to the ladle. The prediction of the temperature in the steel, slag and refractory wall is an important parameter for the refractory loss. We therefore need the temperature in the steel as hot steel is added to the ladle. To account for the heating during the secondary metallurgy we need the electrical power that is added as a function of time. In addition, we need the time history of applied inert gas and the vacuum pressure above the melt. The model must be able to take the entire history of the ladle (since last relining) into account when the simulations are run. The state of the refractory wall from the previous heat, in terms of temperatures and erosion, must be input to the next use of the ladle (next heat). It should be possible to continue the simulation when new data is available.

***Pragmatism Step 2 Analytical Strategy and Plan***

This part included overall model design, resulting in a specification document. The specification document was very detailed, but still only a leading star for the implementation. Instead of writing a specific implementation report the code included necessary comments and the original specification was updated when changes were made. When it comes to the functionalities (how to use the model) no specific documentation was done except short text files to explain the scripts.

The ladle and refractory, with bottom porous plug and taphole is truly 3-dimensional. However, to develop a fast model that can simulate the refractory behaviour over weeks in real-time, using a full 3D model, and with the available time and resources, was deemed infeasible. However, by working with an even simplified 2D model a lot of the features of the ladle can be reproduced. It was therefor decided to move on with a 2D model.

The model will have several simplifications, listed below:

> a) During part of the ladle operation the ladle is inside a ladle furnace (SM mill) where heating electrodes (Ohmic heating) are lowered into the slag. These electrodes have a heating efficiency (tuning coefficient) and dynamic effects of heat storage in the electrodes is neglected. However, at a later staged it was learned that the electrodes do not contact the slag or metal, but are heating with attaching arcs, which jump between electrode tip and slag/metal. The arcs radiate towards the refractory bricks placed above the level of liquid steel and slag. This radiation may be excessive in some cases.

> b) The additions to the ladle will need time to melt and mix. It is assumed that the melting process is instantaneous, as well as the mixing process. As a result, the model will predict an immediate temperature drop due to additions to the melt, while in reality this cooling effect will be distributed over several minutes.

c) Vertical heat conduction between the bricks and inside the steel casing is neglected.

d) Steel and slag temperatures are represented by mass averages.

e) When hot steel is teemed into a colder ladle thermal cracking will occur, increasing with increasing temperature difference. These effects are very hard to model in detail and are proposed to be dealt with as a hybrid extension of the model.

f) Excessive erosion of refractory, above the slag/metal level, happens due to the extremely high temperature of the electrode arcs, together with irregular splashing of hot metal during vacuum treatment. These erosion phenomena, happening above the average melt surface, are not included in the model.

Effects which are dealt with are dynamic temperatures in the side and bottom refractory bricks, insulation layer and steel casing. When stirring gas is applied 2D CFD simulations were performed to compute the distribution of wall shear stresses. Based on a set of these CFD simulations the wall shear stress as function of gas flowrate and relative height could be extracted and then used as input for heat- and mass-transfer models. Based on visual observations from video at the plant, at a late state of the project, it was found that application of the vacuum process, together with gas stirring, led to a violent agitation in the steel close to the surface. This observation resulted in recalculations with CFD, now accounting for the gas expansion due to the local steel pressure. The result was much higher shear stresses close to the surface when vacuum was applied. New fitting functions for the wall shear stress as result of relative height, gas flow rate and pressure above the steel was created and implemented into the model. It is noticed that when other team members visited the plant earlier in the project the implication of vacuum was not realized.

It was assumed that the slag behaved like a moving lid, floating on top of the liquid steel. The modelled wave agitation of the slag, caused by the gas stirring, provided an added local mass transfer rate for refractory dissolution into the slag. The assumption of the slag behaving as a lid should later be relaxed. This will however require more complex and time consuming CFD work. This would improve the model in the slag-metal transition region.

Sub models for refractory dissolution and erosion of the steel wetted refractory, as well as dissolution into the slag, where it is present, had to be developed. Data for solubility of refractory binder in the steel and refractory solubility into the slag, was obtained from literature and from thermodynamic equilibrium software ("FactSage.com," n.d.). The energy equations for slag and steel were written in terms of enthalpy, allowing for any relation between temperature, composition, and enthalpy. This is important when dealing with cold additions to hot slag and steel.

*Pragmatism Step 3 Architecture of the Analytical Framework*

Step 3, the architecture of the analytical framework, designs both the experiments, decides data structures and related analyses, model and simulation entities in greater detail. After this step the development team should be ready for orchestration of the experiments, analyses, models and simulations and data/information exchange among them. Of course, this phase is usually carried in several iterations, usually starting with the proof of the concept model (simplest possible representative model), which gradually approaches the final result, i.e. the framework ready for the execution of the workflows (step 4).

The architecture of the model was created in phases. In the first phase, a simplified proof of the concept model was quickly implemented (as a monolithic approach), to see how the specification was holding up, and if more input was required. This proved valuable, as several issues were handled early. The model was decided to be built in Python.

Once the basic model was working satisfactorily, the implementation was redesigned as a set of modules with well-defined interfaces to give the required flexibility in future applications of the model. We needed a model that could keep its state and be flexible enough to enable changes without major rewrites. The model itself was implemented as a single class, as proved valuable as we had to do several rewrites along to accommodate for unforeseen changes. The use of the model was set up as a series of input scripts, executing runs for the LadleModel object, and with different purposes. For instance, i) running a single case, ii) tune the model with a set of parameters based on one or many cases, and iii) running of entire campaigns from first use till demolition.

The data was originally given as column-based data files (MS Excel and csv). In order to efficiently utilize the data, we had to pre-process them to fit our needs. For instance, the time dependent data was given as large chunks containing multiple heats. These were split into one file for each heat. Later the data was put into a database (InfluxDB), and the data reading had to be changed to accompany two different sources. In hindsight it would have made sense to use a database in the first place. The output of the model was handled as a mix of plots, output to screen and saving of results to file. In the case of the data coming from a database, the results were written back to the database.

Data was originally provided as csv-files. At a later state data was loaded into an InfluxDB[2]. If this had been done at a much earlier stage this could have save time.

*Description of model implementation*
The physics-based model developed here is implemented in Python as a single class. The reason for this is to be able to store a complete state of the model to disk and continue at a later state. This is important as the final version of the model will model several different steel ladles in parallel. Each heat-run of the ladle that should be simulated depends on the previous modelled state. As the simulation time is specified to be significantly shorter than one hour (seconds in reality), while each lade is used approximately twice or three times a day, we need to be able to start and stop the simulation in an easy way.

---

[2] https://www.influxdata.com/

*Table 2. Decision gates in model development workflow / model development process. See also Figure 3.*

| Decision Gate | Outcome | Resulting procedure scenario | Comment |
|---|---|---|---|
| 1. Case definition accepted | yes | Now a physics-based model can be created | |
| | no | Ask industry case owner for more information | |
| 2. Involved physics sufficiently understood? | yes | Implementation can start | Implementation can start as soon as each sub-model has reached a proper understanding and is well-specified. |
| | no | Involve specialists and relevant literature. | The industry case owner can be helpful at this stage |
| 3. Available data understood? | yes | The model can be run | The model can be run as long as the provided data has the right structure. Proper data is only required for decision gate 4 and 5 |
| | no | Go back to the decision gate 2 | |
| 4. Model acceptable as a sub-model in a hybrid model? | yes | Sub-model can now be used in a hybrid model | |
| | no | Go back to the decision gate 3 | This could be due to data not being understood, assumptions in the physics being wrong or the case definition not being properly understood |
| 5. Model can reproduce the data? | yes | Model finished | |
| | no | Go back to the decision gates 2 and 3 | If tuning model parameters is not sufficient to reproduce the data, both the data and the physics should be reviewed |

The main ladle model does depend on several state-less sub-models, all described in (Johansen et al., 2023).

The model is relying on both static and transient data from the plant to run. The data can be retrieved either from an InfluxDB database or from files on the disk. Either way, the time it takes to retrieve the data is relatively time-consuming, and is therefore done only when necessary, and the required data is stored inside the object, and used when needed. When a new heat-run is simulated, a new set of data must be loaded, and the previous data set is overridden. Since the data will be loaded several times for the same object, the data is loaded independently from the model initialization.

Depending on the model scenario, the same model with the same data could be run several times. For instance, the time it takes, from the casting is finished and the ladle is filled with steel again, is modelled as an empty ladle. This is done before the model is run again, but this time the full model and data is used. This way of using the model requires the possibility to reset parts of the ladle state between the different runs. The temperature in the ladle wall, for instance, should not the reset, but things like the amount of steel and simulation time should be reset.

The actual simulation of the refining of the steel ladle is stepping in time with constant time steps. First a preparatory step is done, where the amount of steel and slag in the ladle is determined, the heat added to the ladle during the time step is calculated, and the gas rate and pressure is extracted from the data. In addition, the fraction of steel and slag for each cell is determined, and the mass lost from the refractory during the time step is calculated.

Next the new temperature in the steel and the slag is solved for. With this given, the temperature in the wall and the bottom layer is calculated. Once the model is solved, time dependent data is stored inside the object before the next time step is done.

The mass loss (erosion) of the wall is calculated, and accumulated for each time step, but the wall is only eroded at the end of each heat. Due to the different modes that the model is run in, the actual wall erosion is controlled from the outside as an explicit call to erode the wall. This is done to make sure the temporary runs to create the correct wall temperatures are not affecting the wall thickness, as they should not.

The model should always be used with an external script that sets up a given scenario that should be run. How the scenario is set up with have a large impact on the results.

After a ladle is built with a new refractory lining, the ladle is used many times (40-50) before parts of the refractory are replaced with new refractory. The ladle is then used until all the refractory had to be replaced. Between each use of the ladle, the wall is not allowed to cool down (if the waiting time is too long the refractory wall is heated with burners, this is not part of the model), thus the state of the refractory wall at the end of one heat and the waiting time until it is used again, are both important for the next simulation. All must be taken into account when a simulation is run. This is done by allowing the user to control the model from the outside.

When a ladle object is created, no data is read into the object, which means running the model at this stage would fail. The reason for doing it this way is to avoid having several ways to set the data, and to be able to use the same object for consecutive heats without copying results.

To show how this can be done, we will go through a couple of different scenarios.

First of all it is important to be able to run a single heat independently, and with the possibility to reproduce the results quickly. This way of running will not take the history of the ladle into account properly, and we need a way to estimate a realistic starting state for the refractory wall.

First we have a method to set the initial wall temperature (to not start from a totally unrealistic state, which would take a long simulation time to get to something realistic). This method will give a linear profile between a user given temperature for the inner and outermost bricks. After reading the relevant production data into the case, we can run the case for a given amount of time to heat up the refractory to a realistic temperature. From the casting is finished to the same ladle is used again, the ladle will stay idle for some time. This will cool the refractory significantly. To account for this, we can run the model without steel and slag for a given amount of time. The model will not be able to solve all the equations properly with zero amount of metal, and to avoid problem, a flag is set, telling the model that the simulation is run with an empty ladle. The ladle state is now ready to run the actual simulation. Due to the different modes described above we have an additional method to actually erode the ladle wall at the end of the simulation. This is to make sure that the temporary simulations are not changing the refractory thickness.

A more realistic scenario is to simulate an entire lifetime of a ladle. This can be done in much of the same way as described above. For the first heat, the recipe will be identical, while for the next heats we can use the results of the previous heats as the initial state of the refractory. In this case we can also take into account the actual time between each heat, as the waiting time between heats is recorded. This is now used to calculate how long the ladle is run empty. Sometimes the waiting time is so long that a burner has to be used to keep the refractory wall warm. This is not simulated, and therefore we ignore waiting times longer than three hours in our simulations.

We had an additional challenge; the initial steel temperature in the ladle was unknown. As a compensation for this we received the EAF temperature, but we found that this is not always representative for the starting state. To compensate for the unknown steel temperature, we iterate to find the initial temperature that gives the smallest distance between the calculated steel temperature and the measured temperature.

### *Pragmatism Step 4 Execution*

Step 4 orchestrates workflows of experiments, models and simulations and executes related data analyses. Ideally it should be possible without any framework changes to repeat the exercises and include them as an integral part of the industrial process. However, usually analyses in the evaluation step (step 5 in Figure 1) require repeating the steps 3 - 4 until the framework reaches the quality needed for the support of the industrial process.

The model exemplifies hybrid modelling, where we exploit both static data and dynamic data. Static data includes ladle materials, geometry, last temperature before the metal was transferred to the ladle, time for repair of the refractory and total number of uses at the time of full relining of the ladle. At both relining and demolition (full relining) the erosion profile in the ladle was mapped. Dynamic data include gas purging, vacuum evolution, heater power, steel temperatures (probe based), addition of different alloy and slag former materials, time of tapping, and resting time until next use.

The output data from the process is the measured steel temperatures and the data for relining and demolition. The number of uses before relining and demolition depends on the operator's assessments due to visual inspection. The erosion profiles are maximum values and must be compared with the predictions which are ensemble averages.

The execution step was found to be far from linear as it must involve multiple iterations. Based on initial execution of the model, using available input data, several issues on poor representation of data were found. As we see from Figure 3, decision point 5); when the model fails to reproduce data, we have a vertical arrow back to further calcifications and resulting in updates of the model specification. This process was done many times, all the way until the end of the project.

A good example to show that industrial data not always reflect what you might first think, is the steel temperature data reported by Sidenor. The logging system is reporting a new temperature every second, but from the data we could see that the temperature was constant for a long time, and then suddenly did a jump. We could quickly confirm with the industrial partner that the logging system would repeat the last value (of temperature) entered until there was a new value. In practice the temperature was measured at irregular intervals during the heats. We quickly compensated for this by making a linear interpolation between the measured points. The temperature series is used to compare the calculated values with the measured, but are not used as input to the model, with one exception. The first temperature point was used as a starting value for the steel temperature in the ladle.

As the model improved, and we started running more cases, we realized that the first temperature sometimes seems off. For instance, we found cases where the temperature increased without any energy being added. Further investigation revealed that the first temperature "measurements" for a heat was the last temperature measurement from the previous heat. This meant that we had no value for the critical starting temperature in the ladle. We then asked if we could get temperature measurements from the EAF to use as a starting value. This also proved to be unsatisfactory. We then decided that the best way forward was to iterate on the steel temperature by using the EAF temperature as a starting value, and minimizing R0 (see equation [2]) to a given tolerance, chosen to be 10 K.

For improving the model, we defined a set of tuning parameters. We then simulated the erosion state and temperature of the ladle continuously over many heats, until the maximum erosion of the refractory was 75 %.  This can be compared with the dynamic measured steel temperature in each heat, as well as the number of heats that was run until repair was performed.

Another critical input for the numerical model is the amount of steel in the ladle. This is given in the data, but we found that sometimes the results from the numerical model gave a relatively poor match with the data, and the reported amount of steel seemed off, either too high (more steel than could fit in one ladle), or very low. By going through the steps of pragmatic modelling, we found out that the given amount of steel in the ladle was what was cast, and not a direct measurement of the ladle. It happens that the casting experience issues that result in abortion of the cast. This will lead to a reported steel weight that is less than was actually used in the steel refining. The remaining steel will then end up being registered to one (or several) later heats. There is no way for the numerical model to compensate directly for these errors, like was done for the steel temperature. To avoid too big discrepancies, we limit the minimum and maximum amount of steel added to the ladle.

Tuning parameters that were selected were i) Refractory conductivity, ii) Melting heat for each of the additions, iii) Heat transfer coefficients (external, external emissivity, metal-wall, slag-wall, metal-slag, and slag, refractory and lid emissivity), iv) Electrode energy efficiency, and v) carbon diffusion length in wear bricks. Here only v) deals purely with erosion.

During testing of the model, it was found that the erosion state of the refractory and the evolution of temperatures were closely coupled. In Figure 4 we see that steel temperature is larger than for an eroded refractory compared to an almost uneroded refractory (Figure 5). This is a result of less heat capacity in the eroded refractory. As expected, it was found that when the refractory was cold at the time of filling by steel, the steel temperature is lower and more heating power is needed.

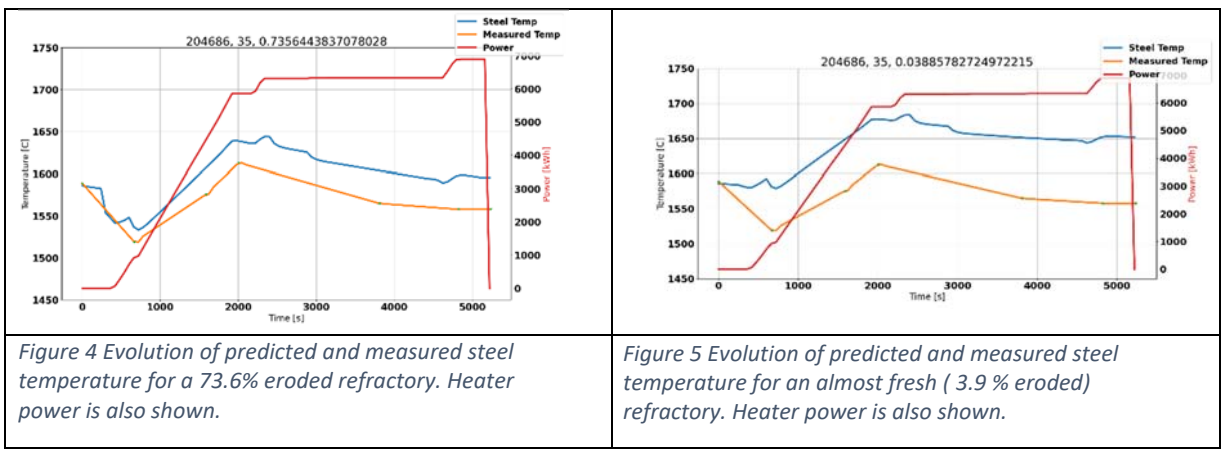

| Figure 4 Evolution of predicted and measured steel temperature for a 73.6% eroded refractory. Heater power is also shown. | Figure 5 Evolution of predicted and measured steel temperature for an almost fresh ( 3.9 % eroded) refractory. Heater power is also shown. |

Temperature tuning was done in two steps, using preliminary and approximate erosion model. As we found that the initial measured temperature in the data was not relevant, we also needed a strategy for obtaining a relevant start temperature for the steel.

*Tuning step a):*

RMS residual for temperature was defined as

$$R_{1,n} = \sqrt{\frac{1}{N_k} \sum_{k=1}^{N_k} \left(T_{k,n}^{pred} - T_{k,n}^{measured}\right)^2} \qquad [1]$$

Here n expresses a campaign number and $N_k$ (k=1, .., $N_k$) is the number of temperature measurements in one heat. Now, if the initial steel temperature is incorrect that will drive a large residual $R_{1,n}$. However, this problem is picked up by the residual $R_{0,n}$, defined as

$$R_{0,n} = \frac{1}{N_k} \sum_{k=1}^{N_k} \left( T_{k,n}^{pred} - T_{k,n}^{measured} \right) \qquad [2]$$

If the predictions are perfect we have, due to incorrect initial temperature for the steel, that $T_{k,n}^{pred} - T_{k,n}^{measured} = \Delta T_{initial}$, resulting in $R_{0,n} = \Delta T_{initial}$, and $R_{1,n} = |\Delta T_{initial}|$. On the other hand, if the temperature levels of both simulation and data are identical, we have that $R_{0,n} = 0.0$, and $R_{1,n} = \sqrt{\frac{1}{N_k} \sum_{k=1}^{N_k} \left( T_{k,n}^{pred} - T_{k,n}^{measured} \right)^2} \neq 0$. Accordingly we decided $\Theta$ as the overall residual to minimize:

$$\Theta_n = \sqrt{R_{1,n}^2 - R_{0,n}^2} \qquad [3]$$

*Tuning step b):*

Here we correct the initial temperatures in order to have the correct steel temperatures for the simulation of refractory erosion. Based one Step a) the initial temperatures are corrected for all cases where $|R_{0,n}| > 10 \text{ K}$.

Note that in both step a) and b) the erosion is predicted, based on preliminary tuning. As the refractory is eroded this will also impact the thermal dynamics of the system.

*Tuning step c):*

Now we tune the erosion part of the model. We have data on when it was decided to repair the refractory and when it was demolished.

We do not have a model for the thermal shock degradation, and this element is for now not considered. As the thermal shock is most important at the bottom of the ladle, while chemical erosion is most pronounced at the slag line, this omission may not be critical for the usability of the model. Accordingly, we tune the erosion part of the model to match the observed times of repair.

Repair is going to happen when the maximum erosion is passing 75 % of the three inner bricks.

We have here a new residual $\Theta_{repair,n}$ to minimize.

$$\Theta_{repair,n} = \left| N_n^{repair,pred} - N_n^{repair,data} \right| \qquad [4]$$

Here N represents the number of heats and n is the campaign number.

*Tuning step d):*

When we can reproduce the times of repair well, we move on to reproduce the use numbers for demolition. The ladle repair means that the bricks above a given level is repaired, while those below are not repaired. This must also be considered for the tuning. Optimally, we should find that the tuning of the demolition is not required. However, it is possible that the repair has changed the properties of the refractory in a way that necessitates some tuning of the models to handle the erosion evolution after repair.

In this case it was eventually found that no tuning behind step c) was necessary. The model could reproduce the demolition data very satisfactory.

### Pragmatism Step 5 Evaluation of the Solution

The outcome of the development is twofold. We have a model that can deliver certain prediction results. In addition, we have a numerical code that can be applied as an element in multiple applications such as cognitive digital twins and other applications for asset management and optimization.

The solution quality is in this case exemplified by a comparison between prediction and measurements at time of demolition, for all ladles and ladle campaigns operated by Sidenor in 2019. The result is shown in Figure 6. The averages are done over bricks 7 – 35, referring to Figure 7. It must be noted that the measurements have picked the bricks which are most eroded at each brick level (referring to brick number in Figure 7). In Figure 7 we see typical measured and predicted erosion profiles. The measured values are a result of the ladle being sectioned in two halves, and where the most eroded bricks for each half are measured. As a result, the prediction model should predict lower values than what is observed. This is also the case as seen in Figure 6. We further see from Figure 7 that high erosion is found above brick 35, "marked splash based erosion". This erosion is a result of thermal shocks due to intermittent splashing of steel due to vacuum treatment, combined with low pressure chemical decomposition (Jansson, 2008) of the MgOC bricks, and where none of these mechanisms are accounted for in the model.

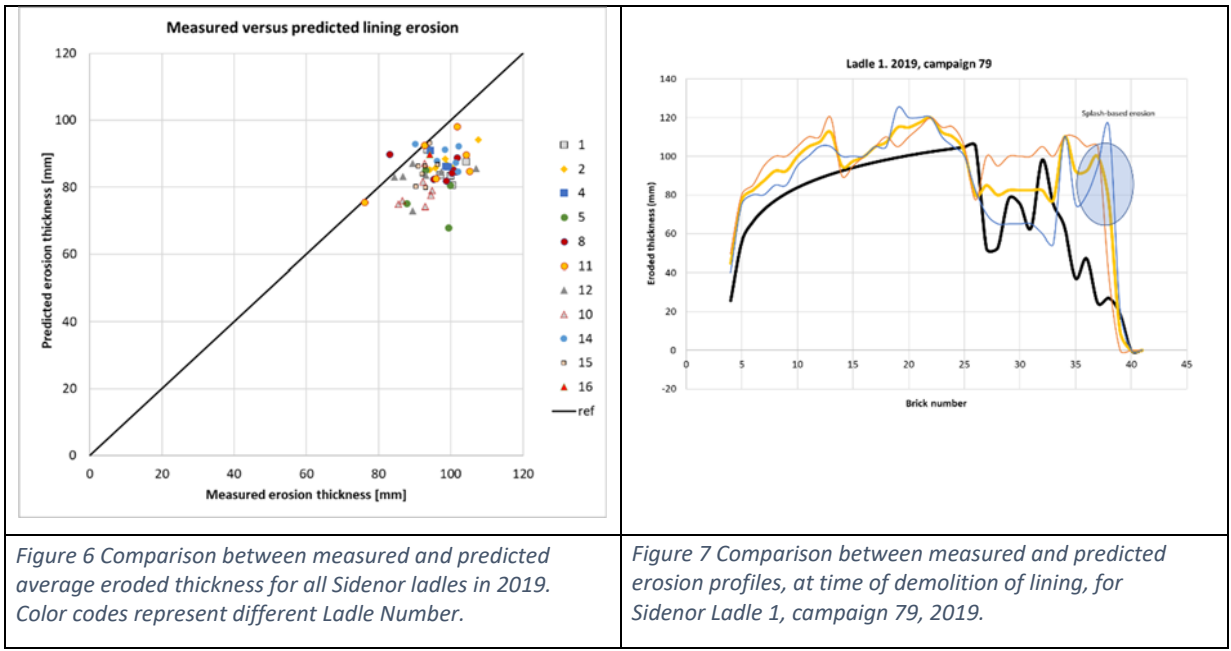

| | |
|---|---|
| Figure 6 Comparison between measured and predicted average eroded thickness for all Sidenor ladles in 2019. Color codes represent different Ladle Number. | Figure 7 Comparison between measured and predicted erosion profiles, at time of demolition of lining, for Sidenor Ladle 1, campaign 79, 2019. |

It is fair to ask if the model can support the operators in allowing more uses of the refractory before demolition? Based on the result in Figure 6 it seems that the answer is yes. The model shows a good comparison between measured demolition data and what is predicted. All campaigns with predicted erosion thickness below 80 mm could be safe to extend with more heats. If the model predicts that erosion is not excessive and the operators is uncertain, this could result in one more heat. We have seen that some heats may have as much as three times or more erosion than average heats. This knowledge would also be useful for the operator's assessment.

*Pragmatism Step 6 Conclusion and Communication*

The conclusions are in this case reserved for the final chapters of this paper. Communications is done internally in the team and with the industry partner. The present paper is an important part of the communication, together with a technical paper (Johansen et al., 2023) that outlines the details of the physics-based simulation model.

**Observations and learning**

The steps in the pragmatism workflow of the presented use case had to be adjusted due to limited time and resources. As seen from Figure 3 multiple feedback loops in the workflow had to be implemented. This was critical to continuously improve the understanding of the ladle process, the data and the involved physics. The work was done with an absolute minimum team. Such a small team is typical for many industrial developments. Therefore, the several learnings from this work may be useful to future developments:

- The overall development would have been faster if the data had been organized in a database in the first place (such as TimescaleDB[3] or InfluxDB). This would have allowed for a more generic pre-processing and presentation of data and would have saved significant time at later stages in the project. At the same time the initial development would have taken more time.
- The code should be modularized as early as possible. This makes the code more versatile to use (testing, tuning, prediction) and easier to develop and later extend. As our model could be accessed as one object, specific scripts could be deployed for the needs at any stage of the development.
- It is very difficult to design a model architecture for the start of the project when so many changes and iterations are needed. It is then expected that several redesigns of the code is required. The initial design should be simple, but effective, and several redesigns should be expected.
- Need of maturing time: The duration of the work should be sufficiently long to allow maturing / better understanding of (a) case, (b) data, and (c) underlying physics. When the model is applied and does not fit with data this most often push the understanding to a higher level.
- More iterations needed than expected: This is in duality with maturing time. For the increased maturing time to make a difference, more iterations in the workflow is a must.
- It was experienced that data available for model tuning was scarce, even if the amount of input data was significant. Data for temperature validation for the slag was not available, and the same was the case for refractory temperature below the steel surface. The only information available is the state of the refractory for the repaired bricks (typically, after more than 40 heats) and the state of the refractory at time of demolition. The erosion difference between heats is only obtained from our model predictions. The model predicts a one-dimensional erosion profile while the data show variation in erosion along the perimeter of the ladle. The details of this variation have recently been recorded. Unfortunately, this information is coming too late to be processed in the COGNITWIN project. This information is critical for a more quantitative assessment of the stochastic variations in erosion which is beyond the capabilities of the current model. Processing this information, to get the variability in erosion at different levels above the ladle bottom, would help to interpret the model predictions in terms of maximum erosion at different levels above the ladle bottom.
- Industrial data is not always what it seems to be for outsiders. Data documentation might sometimes rely on inhouse knowledge that is not transparent for outsiders. Thus is it important to question all data that could not be explained. The data in itself might not be wrong, but the interpretation could be.

---

[3] https://www.timescale.com/

At the end one could ask why not go for a pure ML approach here? This has been attempted but found challenging as the amount of output data is very limited. Such an approach was however explored by (Mutsam et al., 2019) and they obtained acceptable agreement between data, applying both a linear regression model and a deep learning neural network model. As part of pre-processing of their data they removed outliers (unexpected high erosion spots) from their data set.

The difference between our and these approaches is that we have physical mechanisms which we can touch and manipulate and, when tuned to data, it allows us to work outside the data window. This cannot be done safely with models relying only on interpolating data.

After being tuned to data the physics-based model is already a hybrid digital twin. A natural next step is to explore the deviation between the model's predictions and the results obtained by various/alternative ML methods. This could help to single out missing mechanisms, as well as the degree of randomness in the data (from causes we have not recognized or measured).

A final aspect here is the introduction of cognition into this task. This may happen through various mechanisms, such as: i) The operators use the model actively and build experience on how the model predictions and visual observations relate. This will build trust in the model in cases where the operator has doubt about deciding for one more heat. ii) The model predictions, together with operations data, may be presented to the operators as knowledge graphs[4]. This may offer additional support to the operators (Albagli-Kim and Beimel, 2022). iii) Self adaptive algorithms, by learning from data, may continuously improve the model.

The pragmatic modelling approach has two equally important phases: development and exploitation (including use of the models and data in the overall decision support systems and processes). Both phases require a small, but dedicated team of experts (not necessarily more than 2-3 persons). Their engagement should start with the framework development and continue with the exploitation of models and produced results/data. They should also exploit the potential of the framework and the continually produced data for further process optimisations and improvements. This requires continuity of the team and availability of the financial resources for a longer period of time. This requires dedicated strategic management support. If not the value of the work be significantly reduced, if not lost.

There should be a plan for internal training and model adaptation in case the model development is outsourced.

**Conclusions**

The *Pragmatism in industrial modelling methodology* was applied and extended to the development of a model for ladle refractory lifetime prediction. The major contributions to the methodology were

　　i)　　Processes in the metallurgical industry are complex in many dimensions. Operational data will have many challenges and sometimes the data does not express what it seemingly is supposed to express. Therefore, it is critical that the solution architects have some experience with this type of industry to have good communication with the industry experts.

---

[4] https://github.com/SINTEF-9012/SINDIT

ii) Developing a model based on a slim team (core team of two scientists) should be stretched out in time, allowing multiple iterations in the development process. Allocating large funding resources to be consumed over a short time would be costly and would produces less valuable results.

iii) A well-defined tuning strategy was defined. However, exact tuning was not possible as data relevant for operation is monitored and not data that would be good for model tuning and validation. As a result, only approximate tuning is possible. Tuning should assure that all qualitative variations in the data is accommodated for. In this case the model can be used in semi-quantitative manner, where the combinations of model predictions, visual inspections of ladle refractory and operators experience, all together, make the foundation for the decision of the operator to decide that the lining should be demolished or not.

### Acknowlegements

The work in this paper was funded by the H2020 project COGNITWIN (grant number 870130). We thank the COGNITWIN consortium partners who were involved into the Sidenor pilot discussions.
### CRediT author statement

STJ: Conceptualization, Methodology, Writing - Original draft preparation ; BTL: Conceptualization, Methodology, Writing - Original draft preparation; TRD: Resources, Investigation, Writing- Reviewing and Editing; JZ: Conceptualization, Methodology, Writing - Original draft preparation

### Nomenclature

| | |
|---|---|
| AI | Artificial Intelligence |
| BF | Blast Furnace |
| EAF | Electric Arc Furnace (https://en.wikipedia.org/wiki/Electric_arc_furnace) |
| Campaign | The campaign, is given an id number, and for given ladle number, starts with the first use with new lining, and ends with the demolition of the lining. |
| LF | Ladle Furnace |
| ML | Machine Learning |
| MLT | Machine Learning Team |
| SM | Secondary Metallurgy (https://www.britannica.com/technology/steel/Secondary-steelmaking) |
| $T_{n,k}$ | Temperature [K] |

$\Theta_n$         Residual, defined by equation [3]

VD         Vacuum Degasser